\documentclass[conference]{IEEEtran}
\IEEEoverridecommandlockouts

\usepackage{amsmath,amssymb,amsfonts}
\usepackage{algorithmic}
\usepackage{graphicx}

\usepackage{textcomp}
\usepackage{xcolor}
\usepackage{url} 
\usepackage[style=ieee,sorting=none, backend=biber]{biblatex}
\usepackage{amsmath}

\usepackage{amssymb}
\usepackage{braket}

 \usepackage{cancel}
\usepackage{booktabs}   

\begin{document}

\title{Bridging Theory and Practice in Quantum Game Theory: Optimized Implementation of the Battle of the Sexes with Error Mitigation on NISQ Hardware\\
}

\author{%
\IEEEauthorblockN{Germán Díaz Agreda}
\IEEEauthorblockA{\textit{Facultad de Ciencias Naturales, Exactas y de la Educación}\\
Universidad del Cauca\\
Popayán, Colombia\\
germandiaz@unicauca.edu.co}%
\and
\IEEEauthorblockN{Carlos Andres Duran Paredes}
\IEEEauthorblockA{\textit{Facultad de Ciencias Naturales, Exactas y de la Educación}\\
Universidad del Cauca\\
Popayán, Colombia\\
\url{https://orcid.org/0009-0008-3243-7684}}%
\and
\IEEEauthorblockN{Mateo Buenaventura Samboni}
\IEEEauthorblockA{\textit{Facultad de Ciencias Naturales, Exactas y de la Educación}\\
Universidad del Cauca\\
Popayán, Colombia\\
mateobuena@unicauca.edu.co}%
\and
\IEEEauthorblockN{Jhon Alejandro Andrade}
\IEEEauthorblockA{\textit{Facultad de Ciencias Naturales, Exactas y de la Educación}\\
Universidad del Cauca\\
Popayán, Colombia\\
\url{https://orcid.org/0000-0003-1406-3064}}%
\and
\makebox[\textwidth][c]{%
  \begin{tabular}{c}
    Sebastián Cajas Ordoñez\\
    \textit{National Irish Centre for AI (CeADAR)}\\
    \textit{University College Dublin (UCD)}\\
    Dublin, Ireland\\
    \url{https://orcid.org/0000-0003-0579-6178}
  \end{tabular}%
}%
}
\maketitle

\begin{abstract}
Implementing quantum game theory on real hardware is challenging due to noise, decoherence, and limited qubit connectivity, yet such demonstrations are essential to validate theoretical predictions. We present one of the first full experimental realizations of the ``Battle of the Sexes'' game under the Eisert--Wilkens--Lewenstein (EWL) framework on IBM Quantum's \texttt{ibm sherbrooke} superconducting processor. Four quantum strategies (I, H, $R(\pi/4)$, $R(\pi)$) were evaluated across 31 entanglement values $\gamma \in [0, \pi]$ using 2048 shots per configuration, enabling a direct comparison between analytical predictions and hardware execution. To mitigate noise and variability, we introduce a Guided Circuit Mapping (GCM) method that dynamically selects qubit pairs and optimizes routing based on real-time topology and calibration data. The analytical model forecasts up to $108\%$ payoff improvement over the classical equilibrium, and despite hardware-induced deviations, experimental results with GCM preserve the expected payoff trends within $3.5\%$--$12\%$ relative error. These findings show that quantum advantages in strategic coordination can persist under realistic NISQ conditions, providing a pathway toward practical applications of quantum game theory in multi-agent, economic, and distributed decision-making systems.
\end{abstract}


\section{Introduction} \label{sec:intro}

Coordination games represent a fundamental class of strategic interactions where players benefit from aligning their choices despite having conflicting preferences. These scenarios arise frequently in economics, distributed systems, and multi-agent artificial intelligence, where efficient coordination mechanisms are crucial for optimal outcomes \cite{shoham2008multiagent, harsanyi1961rationality,osborne1994course}. Traditional game-theoretic approaches rely on classical probability theory and face inherent limitations in achieving optimal coordination, particularly when communication between players is restricted or costly \cite{farooqui2016game, van2020coordination}

Quantum game theory has undergone significant theoretical development in recent years, incorporating concepts such as quantum superposition and entanglement into the analysis of classical games \cite{Joint_Prob,9805433,kjaergaard_superconducting_2020,fuchs_quantum_2020,kairon_noisy_2020}. This emerging field promises to overcome classical limitations by providing players with expanded strategy spaces and novel correlation mechanisms unavailable in classical settings. Among existing problems, the "Battle of the Sexes" (BoS) is a classical coordination game with multiple Nash equilibria, where one can analyze how conflict and uncertainty affect each player's payoff. Its roots trace back to the foundational work of John von Neumann and Oskar Morgenstern in Theory of Games and Economic Behavior (1944), which established the mathematical framework for analyzing strategic interactions \cite{Neumann_1944}. The game was later formalized by Anatol Rapoport and Melvin Guyer as a type of 2×2 coordination game that captures real-life dilemmas where mutual benefit is possible despite conflicting individual preferences \cite{Rapoport_1967}. A common example portrays a couple choosing between attending an opera or a football game, each has a different favorite but both prefer being together over going alone. In the late 20th and early 21st centuries, the Battle of the Sexes attracted interest from researchers in quantum game theory, a field that explores how quantum mechanics can influence and enrich classical game dynamics. Eisert, Wilkens, and Lewenstein (1999) first demonstrated how quantum entanglement could alter outcomes in games like the Prisoner's Dilemma \cite{eisert_quantum_1999}, paving the way for similar quantum adaptations of coordination games. Later studies by Marinatto and Weber (2000) extended these ideas directly to the Battle of the Sexes, showing that players sharing an entangled quantum state could reach fairer or more efficient equilibria compared to classical strategies \cite{marinatto_quantum_2000, ,leal_pareto-optimal_2020,khan_quantum_2025}. These quantum versions allow for richer strategy sets and exploit phenomena like superposition and entanglement, fundamentally challenging the assumptions and limits of classical rationality.

The theoretical promise of quantum game theory, however, faces significant practical challenges when implemented on current Noisy Intermediate-Scale Quantum (NISQ) devices. While quantum non-locality has been successfully demonstrated through experiments such as the Clauser-Horne-Shimony-Holt (CHSH) inequality violations \cite{jaffali_two_2024,jusseau2024fourqubitchshgames,kelleher_implementing_2024,sheffer_playing_2022,waring_chsh_2025}, these fundamental physics demonstrations typically focus on binary outcomes rather than the continuous strategic payoffs required for game theory applications. The transition from theoretical quantum advantages to practical implementations faces the reality of gate errors, decoherence, and limited coherence times that characterize NISQ hardware \cite{bharti2022noisy}.

On the other hand, the CHSH game, a cornerstone in the verification of quantum non-locality, provides an appropriate framework for studying coordination between quantum agents without direct communication, an aspect relevant in the quantum BoS game, where two players seek to coordinate while preferring different outcomes. By incorporating entanglement and superposition, more equitable and efficient strategies are enabled, inspired by the non-local correlations inherent to CHSH. From this perspective, the implementation of circuits based on the CHSH inequality has become a key tool for exploring and validating non-locality phenomena in modern quantum devices. It has been shown that, despite noise and decoherence, significant violations of the inequality can be observed through dynamic circuits and adaptive error correction \cite{jaffali_two_2024,jusseau2024fourqubitchshgames,kelleher_implementing_2024,sheffer_playing_2022,waring_chsh_2025}. These realizations confirm the feasibility of carrying out fundamental experiments on NISQ platforms, while also opening new avenues for the development of more robust quantum communication and cryptographic protocols \cite{lopez-incera_encrypt_2020,sekatski_device-independent_2021,aharon_device-independent_2016}.

Despite these advances in quantum non-locality demonstrations, a critical gap remains between theoretical quantum game theory and its experimental validation on real quantum hardware. Previous implementations have been limited to classical simulations or idealized theoretical analyses, leaving open fundamental questions about whether quantum strategic advantages can survive the noise and operational constraints of current quantum processors. This gap is particularly pronounced for coordination games like Battle of the Sexes, where the quantum advantage depends on maintaining coherent superposition states throughout the strategic interaction process.

In this work, we compare two execution environments of the quantum version of BoS: analytical calculation and real hardware execution on IBM Quantum\footnote{IBM Quantum hardware access was provided through the IBM Quantum Network: \url{https://quantum.ibm.com}}. To do this, circuits were implemented exploring different configurations of the initial entanglement parameter $\gamma$ and various combinations of local gates (Hadamard and $R_y$ rotations) for both players, aiming to maximize their payoffs. In this context, Consuelo‐Leal et al. present, in a purely theoretical framework, a strategy in which no player's payoff can be increased without decreasing the other's \cite{leal_pareto-optimal_2020}, and Romero-Álvarez et al. propose a circuit scheduler to optimize task allocation on cloud QPUs \cite{romero-alvarez_quantum_2024}. While the previous work served as a starting point for designing a circuit-combination strategy, the present study focuses on a novel noise mitigation proposal aimed at optimizing quantum hardware use. This proposal consists of dynamically selecting qubit pairs with the least physical interference and iteratively reassigning them in each circuit, adapting to the specific conditions of the device. Specifically, we introduce a Guided Circuit Mapping (GCM) strategy that addresses the practical challenges of multi-circuit execution on NISQ devices by optimizing qubit allocation based on real-time calibration data and hardware topology constraints. The results show that the proposed mitigation strategy significantly improves the fidelity of quantum payoffs in a NISQ environment.

Our contributions are threefold: (1) we provide the first comprehensive experimental validation of the Eisert-Wilkens-Lewenstein quantum Battle of the Sexes on superconducting quantum hardware, (2) we quantify the performance gap between theoretical predictions and NISQ implementation across multiple quantum strategies, and (3) we demonstrate that targeted error mitigation can preserve quantum strategic advantages under realistic noise conditions. These results establish a foundation for practical quantum game theory applications and provide insights into the requirements for quantum advantage in strategic scenarios.

Section 2 introduces the formalism of the quantum BoS game and details the EWL quantization scheme along with the employed strategies. Next, Section 3 describes the complete methodology, from circuit construction in Qiskit to execution on real quantum devices. Section 4 presents and analyzes the comparative results obtained analytically and on hardware. Section 5 discusses the results, Section 6 summarizes the main conclusions, and Section 7 suggests possible directions for future research.
\vspace{-0.2em}
\section{Theoretical Framework}

The BoS scenario is a classical coordination game between two players who have different preferences but a common interest in acting together. Traditionally, it is illustrated with two individuals, Alice and Bob, who wish to spend the evening together, although one prefers to go to the opera and the other to watch television. The typical payoff matrix for this game can be represented as:


\begin{table}[htbp]
\centering
\caption{Payoff Matrix of the Battle of the Sexes Game}
\label{tab:matriz_pagos}
\begin{tabular}{lcc}
\toprule
\multicolumn{1}{c}{} & Opera & Television \\   
\cmidrule{2-3}                       
Opera       & (3, 2) & (0, 0) \\
Television  & (0, 0) & (2, 3) \\
\bottomrule
\end{tabular}
\end{table}

where the first value in each cell corresponds to Alice's payoff and the second to Bob's. There are two pure Nash equilibria:
\begin{itemize}
    \item Both choose \textit{Opera}
    \item Both choose \textit{Television}
\end{itemize}

each favoring one player more. There is also a mixed equilibrium where both players use a randomized choice strategy based on equilibrium probabilities, although it is suboptimal in terms of coordination.

Lack of communication or common knowledge leads to a coordination dilemma: if they choose different strategies, they get the worst possible payoff $(0,0)$. This structure makes the game a paradigmatic case of coordination with multiple equilibria and compromised efficiency \cite{khan_quantum_2025}.

To maximize both players' payoffs, it is necessary to reach the mixed equilibrium. This fundamental concept in game theory extends Nash equilibria to situations where players choose strategies with some randomness rather than pure strategies. Thus, to find the mixed equilibrium:

Alice adjusts $P$ so that Bob is indifferent between his options:

Bob's expected payoff if he chooses Opera equals the payoff if he chooses TV
\begin{align*}
3P + 0(1 - P) &= 0P + 3(1 - P) \Rightarrow P = \frac{3}{5}
\end{align*}

Similarly, Bob adjusts $Q$ so that Alice is indifferent:
\begin{align*}
3Q + 0(1 - Q) &= 0Q + 2(1 - Q) \Rightarrow Q = \frac{2}{5}
\end{align*}

In this way, neither player can guess the other's strategy. The expected payoffs are given by the following expression:
\begin{equation}
E = \sum (\text{Combined probability})(\text{Payoff})
\end{equation}

Using equation (1):
\begin{align*}
E_A &= 3\left(\frac{3}{5}\right)\left(\frac{2}{5}\right) + 2\left(\frac{2}{5}\right)\left(\frac{3}{5}\right) = 1.2 \\
E_B &= 2\left(\frac{3}{5}\right)\left(\frac{2}{5}\right) + 3\left(\frac{2}{5}\right)\left(\frac{3}{5}\right) = 1.2
\end{align*}

There is also a $48\%$ risk of not coordinating (result $(0,0)$).

An extension of this type of game to the quantum domain was proposed by Eisert, Wilkens, and Lewenstein (EWL) \cite{eisert_quantum_1999}, who introduced a formalism to quantize non-cooperative games. In their approach:

\begin{itemize}
    \item Players no longer choose classical actions, but unitary operators applied to qubits
    \item The game starts from an entangled quantum state:
    \begin{equation}
        \ket{\psi_{\text{in}}} = J \ket{00} = \cos\left(\frac{\gamma}{2}\right) \ket{00} + i\sin\left(\frac{\gamma}{2}\right) \ket{11}
    \end{equation}
    where $J$ is the entangling operator and $\gamma$ controls the degree of entanglement.
\end{itemize}

\begin{figure}[ht]
    \centering
    \includegraphics[scale=0.47]{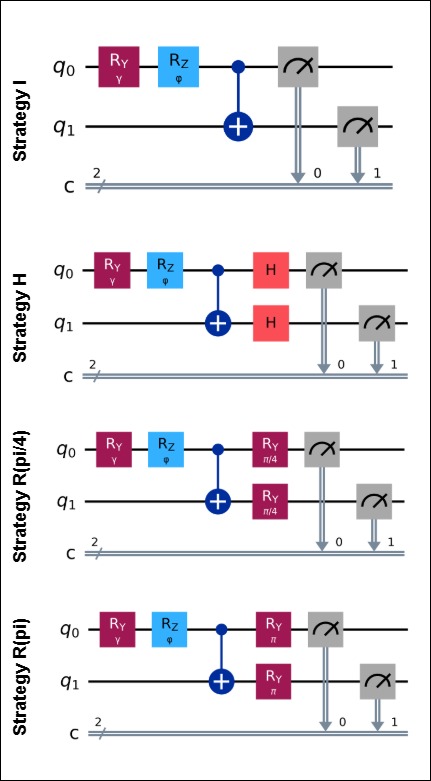}
    \vspace{-1em}
    \caption{\color{black}%
    Quantum circuits used to implement the four strategies in the quantized Battle of the Sexes game.}
    \label{fig:4_Circuitos}
    \vspace{1em}
\end{figure}

The quantum strategies are implemented by applying local unitary gates to a pair of previously entangled qubits, so that each player operates on their qubit with the corresponding gate before the final measurement. As shown in Figure \ref{fig:4_Circuitos}, each quantum strategy consists of preparing an entangled state using parameterized gates, followed by local operations representing the players' choices. The entanglement preparation employs $R_y(\gamma)$ and $R_z(\phi)$ gates, together with an entangling gate (ECR or CNOT), to create the initial state that enables quantum advantages over classical strategies. In this configuration, $\phi$ is set to $0$, as no phase adjustment is applied. Specifically, the Identity gate $I$ does not modify the initial entangled state; the Hadamard gate $H$, $\ket{+} = \frac{\ket{0} + \ket{1}}{\sqrt{2}}$ and $\ket{-} = \frac{\ket{0} - \ket{1}}{\sqrt{2}}$; the rotation $R_y\left(\frac{\pi}{4}\right)$ approximates a $\frac{\pi}{4}$ rotation on the Bloch sphere along the $y$ axis, shifting the intermediate probability amplitudes; and the rotation $R_y(\pi)$ is equivalent to a 180º rotation on the same axis, swapping $\ket{0}$ and $\ket{1}$. After preparing the entangled state using the $R_y(\gamma)$ and $R_z(\phi)$ gates followed by an entangling gate (ECR or CNOT), each player applies their local unitary $I, H, R_y\left(\frac{\pi}{4}\right), R_y(\pi)$, finally, both qubits are measured in the computational basis to assign the payoffs. This ``local gates on entangled qubits'' scheme was proposed in the Eisert--Wilkens--Lewenstein formulation of quantum games \cite{eisert_quantum_1999}. Figure \ref{fig:4_Circuitos} outlines the basic parameterized quantum circuit along with each strategy used in this work.

All circuits begin with entanglement preparation via the gates $R_y(\gamma)$ and $R_z(\phi)$ followed by a controlled-X (CNOT or ECR) gate.
    Each strategy applies a different set of local unitary operations to the entangled qubits:
    \textbf{(Top)} Strategy I leaves the state unchanged;
    \textbf{(Second)} Strategy H applies Hadamard gates to both qubits;
    \textbf{(Third)} Strategy $R(\pi/4)$ applies $R_y(\pi/4)$ rotations;
    \textbf{(Bottom)} Strategy $R(\pi)$ applies full $R_y(\pi)$ flips.
    All circuits conclude with measurement in the computational basis to estimate output distributions.
    These distributions are later mapped to payoffs for Alice and Bob depending on the resulting state $|00\rangle$, $|01\rangle$, $|10\rangle$, or $|11\rangle$.

The general expression of each strategy applied to the entangled qubits (2) is given by (3):
\begin{equation}
    U \ket{00} = U \ket{E}
\end{equation}

Where $U$ represents the strategy and $\ket{E}$ the already entangled state.

After applying $U$ on the entangled state, correlated and anti-correlated states are obtained \cite{leal_pareto-optimal_2020}.
\begin{equation}
    U \ket{00} = P_{00} \ket{00} + P_{10} \ket{10} + P_{01} \ket{01} + P_{11} \ket{11}
\end{equation}

Where anti-correlated states $P_{10} \ket{10}$ and $P_{01} \ket{01}$ are omitted in the payoff as they represent non-coordination states $(0,0)$.

From the functions obtained in (4), the probabilities $P_{00}$ and $P_{11}$ are extracted as $\gamma$ dependent functions, from which expected payoffs are calculated using the following formulas representing the payoffs for Alice and Bob:
\begin{equation}
    E_A = 3\left| P_{00} \right|^2 + 2\left| P_{11} \right|^2
\end{equation}
\begin{equation}
    E_B = 2\left| P_{00} \right|^2 + 3\left| P_{11} \right|^2
\end{equation}

After calculating the probability functions for each strategy, the following equations were obtained for Bob and Alice's payoffs.

\textbf{For strategy I.}
\begin{equation}
    E_{A,I} = 3 \cos^2\left(\frac{\gamma}{2}\right) + 2 \sin^2\left(\frac{\gamma}{2}\right)
\end{equation}
\begin{equation}
    E_{B,I} = 2 \cos^2\left(\frac{\gamma}{2}\right) + 3 \sin^2\left(\frac{\gamma}{2}\right)
\end{equation}

\textbf{For strategy H.}
\begin{equation}
    E_{A,H} = \frac{5}{4} \left[ \cos\left( \frac{\gamma}{2} \right) + 2 \sin\left( \frac{\gamma}{2} \right) \right]^2
\end{equation}
\begin{equation}
    E_{B,H} = \frac{5}{4} \left[ \cos\left( \frac{\gamma}{2} \right) + \sin\left( \frac{\gamma}{2} \right) \right]^2
\end{equation}

\textbf{For strategy $R\left( \frac{\pi}{4} \right)$.}
\begin{equation}
\begin{split}
E_{A, R\left( \frac{\pi}{4} \right)}=\ & 3 \left[ 0.853 \cos\left( \frac{\gamma}{2} \right) 
+ 0.146 \sin\left( \frac{\gamma}{2} \right) \right]^2 \\
& + 2 \left[ 0.853 \sin\left( \frac{\gamma}{2} \right) 
+ 0.146 \cos\left( \frac{\gamma}{2} \right) \right]^2
\end{split}
\end{equation}

\begin{equation}
\begin{split}
E_{B, R\left( \frac{\pi}{4} \right)} =\ & 2 \left[ 0.853 \cos\left( \frac{\gamma}{2} \right) 
+ 0.146 \sin\left( \frac{\gamma}{2} \right) \right]^2 \\
& + 3 \left[ 0.853 \sin\left( \frac{\gamma}{2} \right) 
+ 0.146 \cos\left( \frac{\gamma}{2} \right) \right]^2
\end{split}
\end{equation}

\textbf{For strategy $R(\pi)$.}
\begin{equation}
    E_{A, R(\pi)} = 2 \cos^2\left( \frac{\gamma}{2} \right) + 3 \sin^2\left( \frac{\gamma}{2} \right)
\end{equation}
\begin{equation}
    E_{B, R(\pi)} = 3 \cos^2\left( \frac{\gamma}{2} \right) + 2 \sin^2\left( \frac{\gamma}{2} \right)
\end{equation}

Thus, Alice and Bob's payoffs were obtained analytically. It is important to mention that for each strategy, an equilibrium point was found where the payoffs for the players are equal and optimal, with a value of 2.5 for each strategy. This payoff exceeds the classical payoff (1.2) by 108\,\%.

\section{Method}

This study adopts a mixed analytical-experimental approach, beginning with the formulation of the BoS game in its classical version. A probabilistic approach is employed to determine the mixed Nash equilibrium, by computing the optimal probabilities that ensure no player has incentives to change their strategy.

\vspace{-1em}  
\begin{figure}[htbp]
  \centering
  \includegraphics[width=\linewidth]{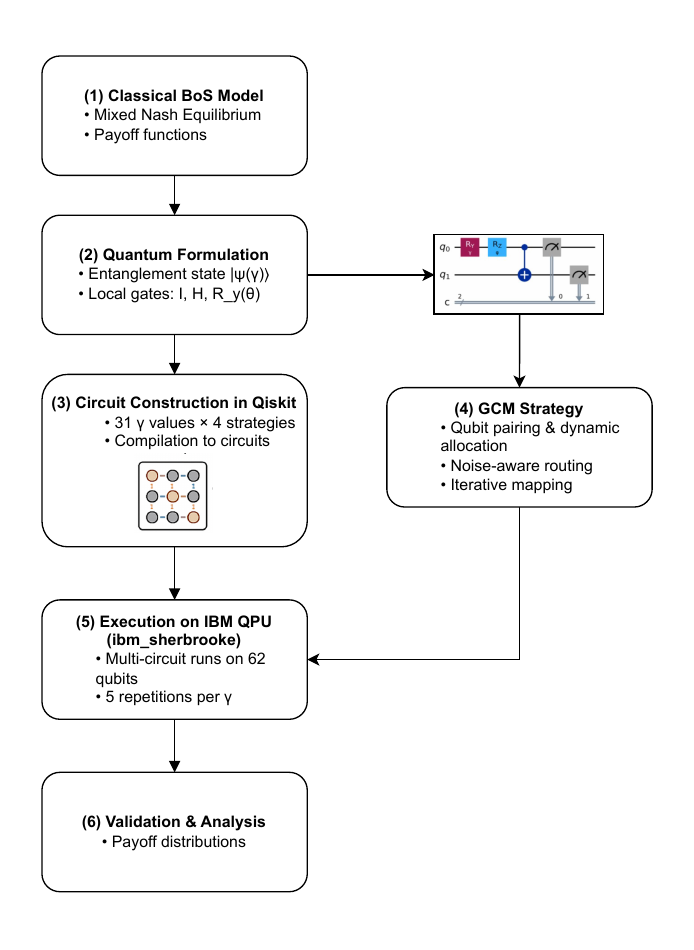}
  \caption{\color{black}
    Overall workflow for implementing and validating the quantum Battle of the Sexes game, from classical modeling to hardware execution and analysis. The feedback arrow indicates how GCM refines mappings based on prior results.
  }
  \label{fig:esquema}
\end{figure}

\subsection{Quantum Model}\label{AA}

Subsequently, the model is translated into the quantum domain through the introduction of entangled states whose amplitude depends on the parameter $\gamma$. Hadamard gates and unitary rotations $R_y(\gamma)$ are applied to the entangled state, exploring variations in the distribution of outcomes and payoffs. This allows for a direct comparison between classical solutions and those obtained by leveraging the properties of quantum mechanics.

As shown in Fig.~\ref{fig:esquema}, the process begins with modeling the classical Battle of the Sexes game, including computation of the mixed Nash equilibrium. Then, entanglement is encoded via the state $\ket{\psi(\gamma)}$, and four local quantum strategies are defined: $I$, $H$, $R_y(\pi/4)$, and $R_y(\pi)$.

\subsection{Experimental Implementation}

The experimental implementation is carried out by connecting to IBM Quantum through the Qiskit environment. All analytical results are precomputed and integrated into the code. A specific class is developed to construct the circuits, execute them, and aggregate the obtained results.
A central component of the workflow is the Guided Circuit Mapping (GCM) strategy, which dynamically selects pairs of qubits with low error rates, performs noise-aware routing, and iteratively adjusts the mapping based on previous runs. This strategy helps mitigate the effects of noise in NISQ (Noisy Intermediate-Scale Quantum) devices.

The raw counts generated by the quantum processor are processed to estimate payoff distributions and compared with the analytical reference. This validates the model’s fidelity and quantifies the deviation across strategies.

\subsection{Execution on QPU}

To execute the experiment in a real environment, the \texttt{ibm sherbrooke} processor was used, connected via Qiskit-IBM-Runtime v0.39.0 \footnote{IBM Quantum hardware access was provided through the IBM Quantum Network. The \texttt{ibm sherbrooke} processor specifications and calibration data are available at \url{https://quantum.ibm.com/services/resources}}. This chip features 127 qubits based on the Eagle r3 architecture, with a minimum two-qubit gate error of approximately $2.5 \times 10^{-3}$, a performance of 150K CLOPS, and average coherence times of $T_1 \approx 286\,\text{s}$ and $T_2 \approx 226\,\text{s}$.

For this study, 31 values of the parameter $\gamma$ are considered. Although each one would typically be executed individually (as shown in Fig.~\ref{fig:4_Circuitos}), the objective here is to perform all executions simultaneously. This implies the independent execution of 31 circuits within a single job, using 62 qubits from the processor.
Given the nature of NISQ devices, a high level of noise is expected, with considerable probability that the results will be masked by the inherent interference of quantum noise.

\subsection{GCM Strategy}

To maximize fidelity in a multicircuit environment, 31 pairs of physically connected qubits were selected from the processor's architecture. Each pair was placed at least one qubit apart from the others to avoid interference between circuits. Simultaneously, a mapping of the architecture was defined to guide the subsequent steps.

Based on the multicircuit design implemented in Qiskit, this mapping was used to assign each schematic circuit to its corresponding physical pair on the QPU, leveraging Qiskit's transpilation parameters. For each value of $\gamma$, the schematic circuit was routed to its mapped physical pair, iterating through the entire list of values. This process resulted in a total of 31 jobs submitted to IBM Quantum.

From the marginal counts obtained for each state, distributed means and variances were computed using specific functions. Based on these results, the effective payoffs and their respective uncertainties were derived, along with miscoordination and the associated error propagation.

This approach enabled a direct quantitative analysis of the discrepancy between the analytical model and the hardware results, revealing the impact of physical noise on the game’s payoffs. As a result, the GCM Strategy is proposed, which involves dynamically leveraging the quantum hardware architecture along with an error reduction scheme.

\begin{figure}[ht]
    \centering
    \includegraphics[scale=0.85]{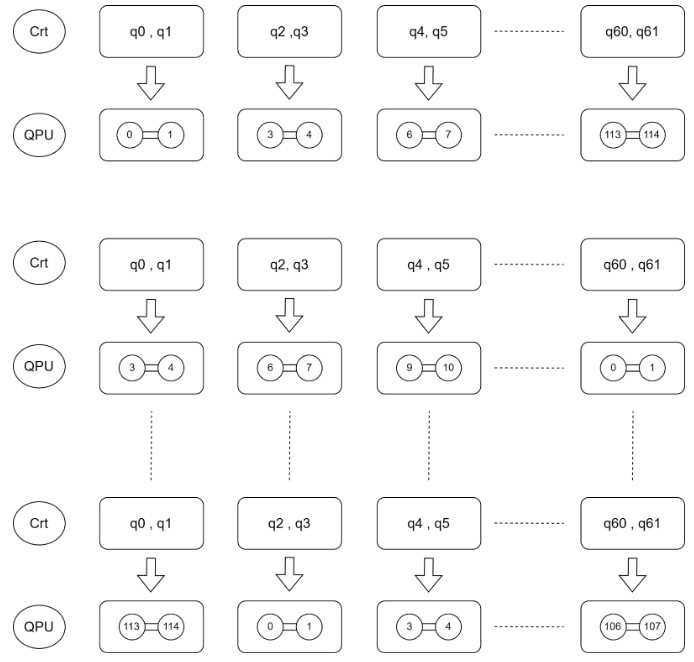}
    \caption{GCM Strategy.}
    \label{fig:strategy}
\end{figure}

This strategy is based on understanding the quantum architecture, performing efficient mapping procedures, and using statistical techniques, as illustrated in Fig.~\ref{fig:strategy}. The entire process was applied to each quantum strategy evaluated.

\subsection{Validation Method}

To verify the robustness of the experimental results, a systematic statistical procedure was applied:
\begin{itemize}
    \item Data Processing: The probabilities of occurrence of the states $\ket{00}$, $\ket{01}$, $\ket{10}$ and $\ket{11}$ were extracted from the raw counts for each of the five repetitions per configuration.
    \item Statistical Analysis: Distributed averages and sample variances were computed using grouping functions by execution index. The effective payoffs for Alice and Bob were derived by applying the corresponding payoff matrices.
    \item Error Metrics: The Root Mean Squared Error (RMSE) between the experimental payoffs and theoretical predictions was computed:
    \begin{equation}
    RMSE = \sqrt{\sum_{i=1}^{n}\frac{(\hat{y_i} - y_i)^2}{n}}
    \end{equation}
    \item Confidence Intervals: 95\% confidence intervals were calculated for each of the five iterations using the Student’s t-test.
\end{itemize}
This approach enables a direct quantitative analysis of the discrepancy between the analytical model and the hardware results, revealing the specific impact of physical noise on the quantum game’s payoffs.

\section{Results}

\subsection{Execution on QPU}

\begin{figure}[ht]
    \centering
    \includegraphics[scale=0.5]{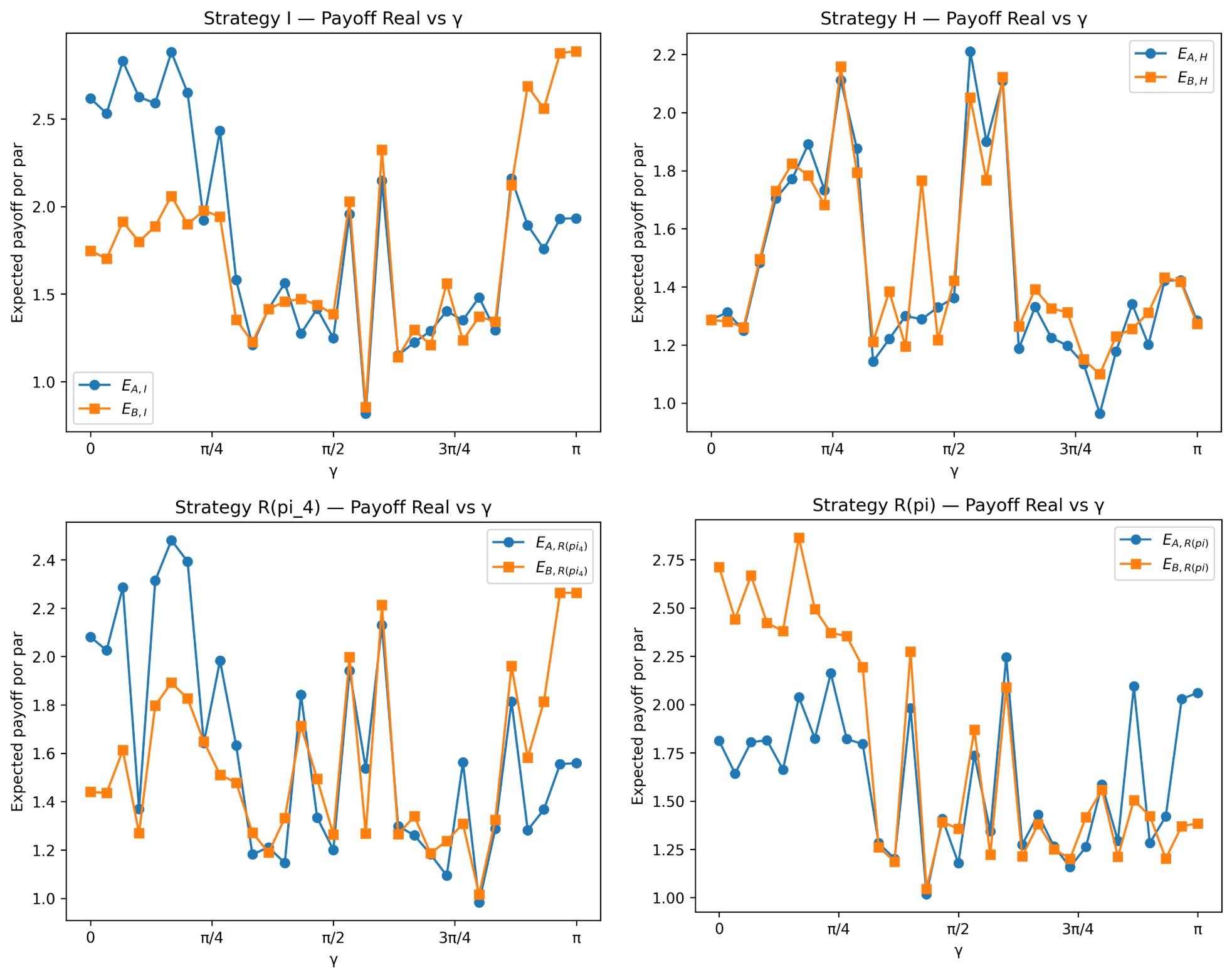}
    \caption{Execution results on QPU without the GCM\_Strategy.}
    \label{fig:juntas}
    
\end{figure}

In the real execution on the \textit{ibm sherbrooke} processor without the use of \textit{GCM Strategy}, the observed payoffs exhibit fluctuations due to physical noise and variations in the fidelity of the selected qubit pairs.

As shown in Fig.~\ref{fig:juntas}, under Strategy I the payoffs for Alice (${E_A}$) and Bob (${E_B}$) show high dispersion and lack of a clear trend, reflecting sensitivity to errors, interference, and decoherence as the number of qubits in use increases. With the other strategies, which involve adding gates to the entangled circuit, the result is not better and resembles random values.

After fully applying \textit{GCM Strategy}, the following results were obtained:

\begin{figure}[ht]
    \centering
    \includegraphics[scale=0.5]{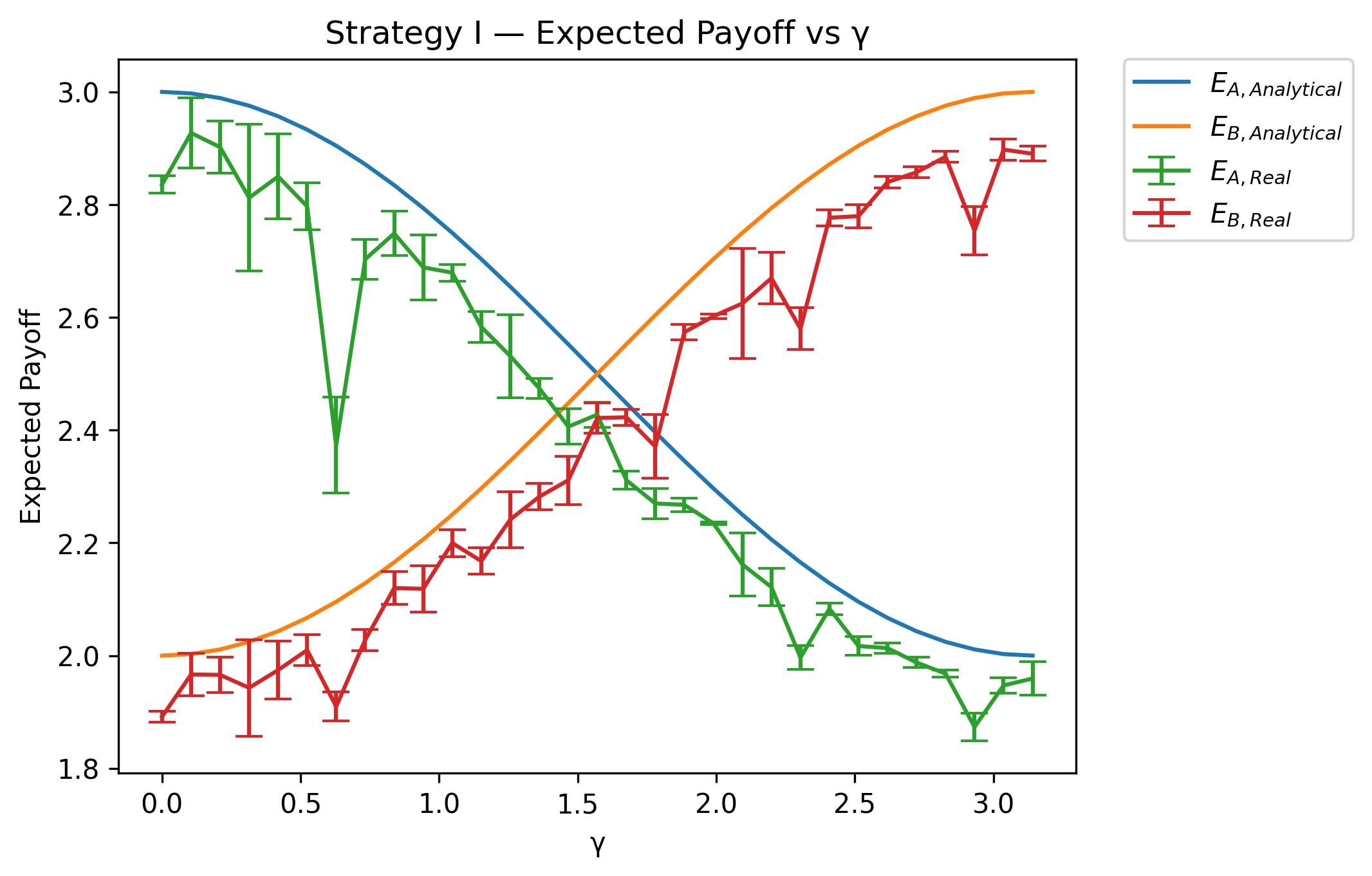}
    \caption{Quantum Battle of the Sexes – Real Payoffs vs $\gamma$ for Strategy I.}
    \label{fig:estrategiai}
\end{figure}

Figure~\ref{fig:estrategiai} shows the results of Strategy I after applying \textit{GCM Strategy}. For each of the 31 values of $\gamma$, five repetitions were executed on the same day. From these data, averages were calculated and plotted with a 95\% confidence interval. In this configuration, the resulting payoff values exhibit a trend that follows the general shape of the theoretical model ($E_{A,Analytical}$ and $E_{B,Analytical}$), albeit with a negative deviation from it.

\begin{figure}[ht]
    \centering
    \includegraphics[scale=0.5]{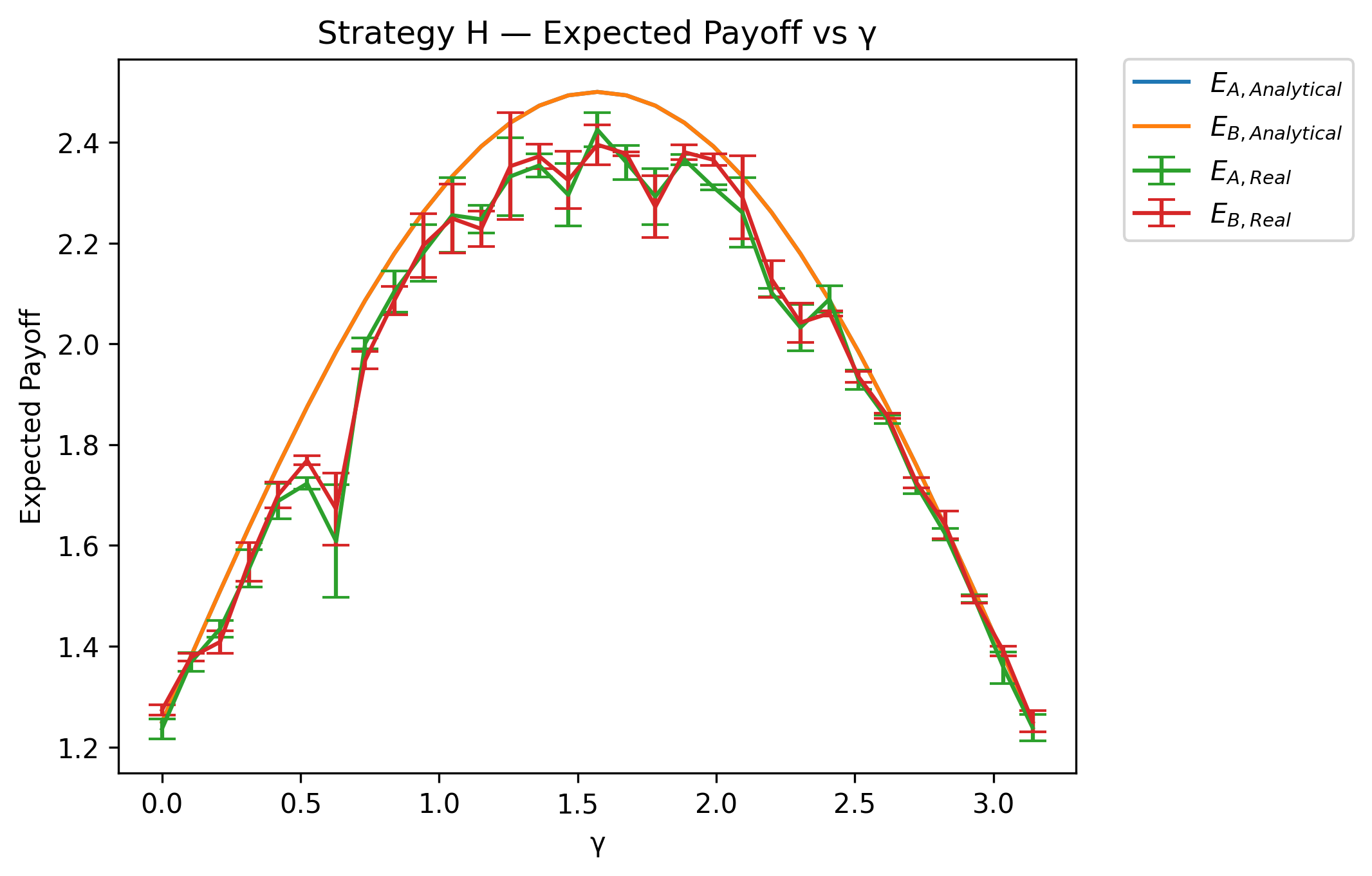}
    \caption{Quantum Battle of the Sexes - Real Payoffs vs $\gamma$ for Strategy H.}
    \label{fig:estrategiaH}
\end{figure}

Figure~\ref{fig:estrategiaH} shows the corresponding results for Strategy H. Comparing the obtained averages, a smaller difference is observed between the experimental payoffs and the analytical values, with a more bounded distribution and less underestimation compared to other strategies.

\begin{figure}[ht]
    \centering
    \includegraphics[scale=0.5]{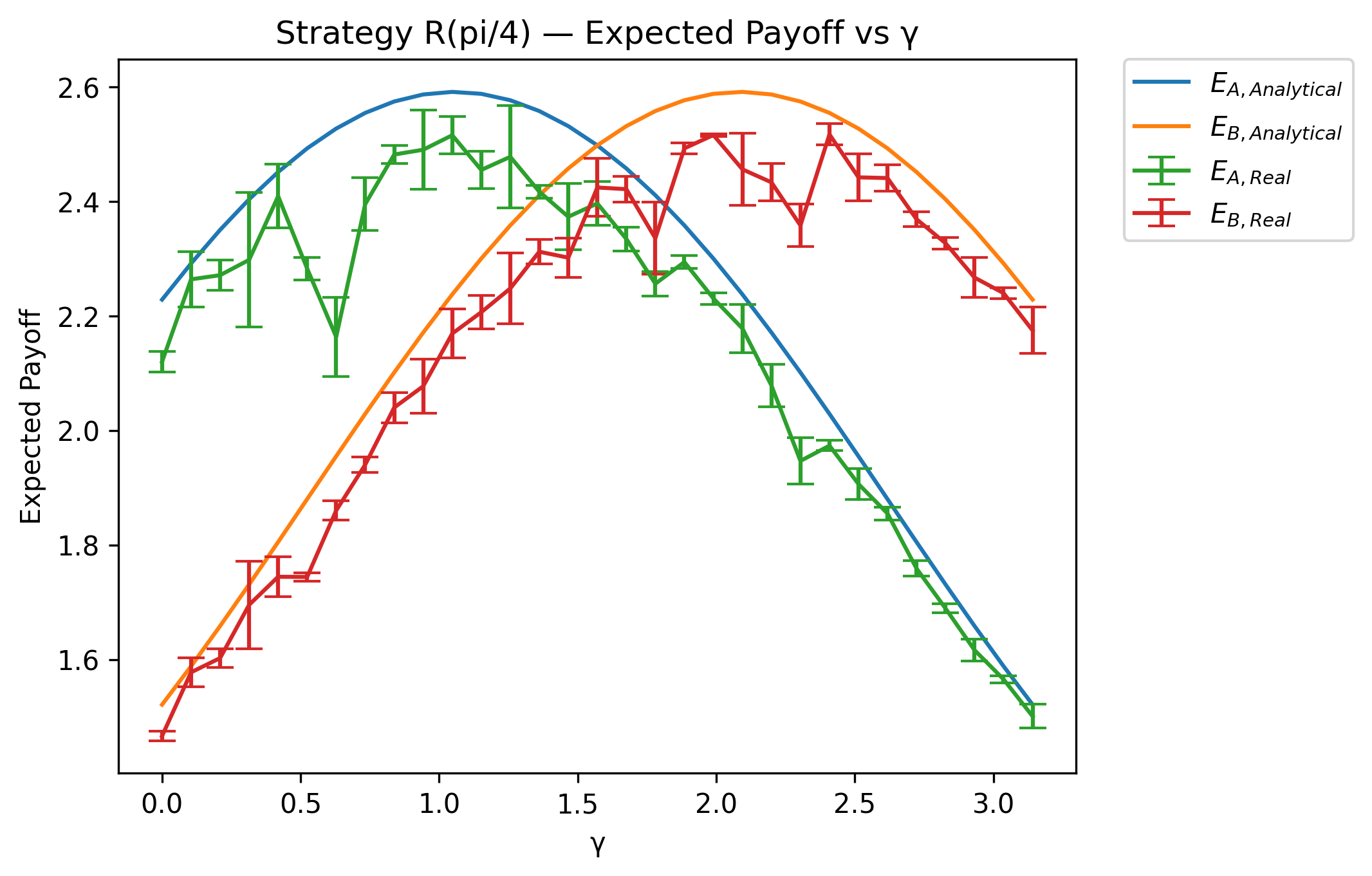}
    \caption{Quantum Battle of the Sexes - Real Payoffs vs $\gamma$ for Strategy $R(\frac{\pi}{4})$.}
    \label{fig:estrategiapi4}
\end{figure}

\begin{figure}[ht]
    \centering
    \includegraphics[scale=0.5]{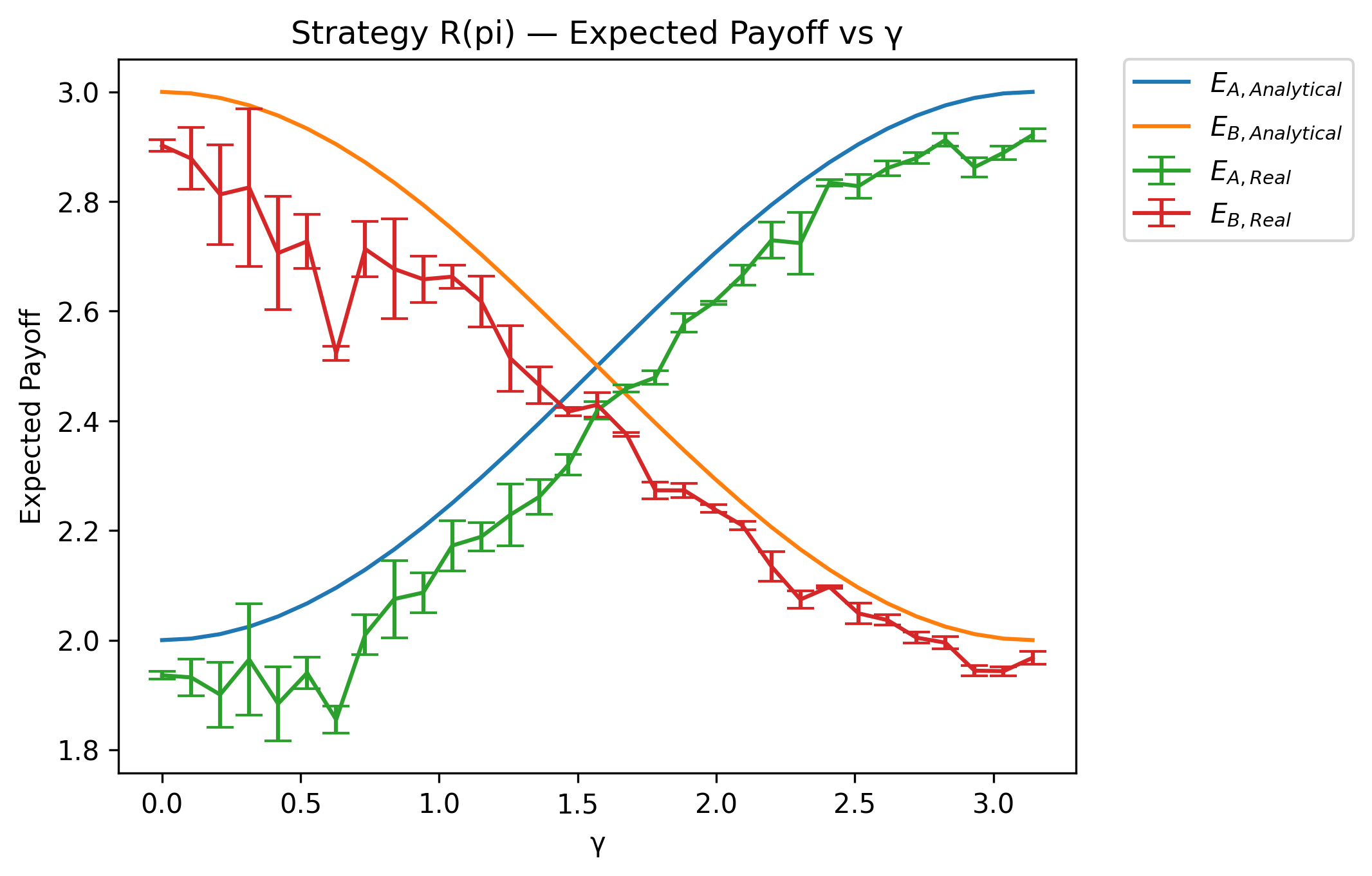}
    \caption{Quantum Battle of the Sexes - Real Payoffs vs $\gamma$ for Strategy $R(\pi)$.}
    \label{fig:estrategiapi}
\end{figure}

The graphs in Figs.~\ref{fig:estrategiapi4} and \ref{fig:estrategiapi} correspond to the strategies $R\left(\frac{\pi}{4}\right)$ and $R(\pi)$, respectively. In both cases, the average payoffs for each $\gamma$ value retain the expected curve shape, although they show greater deviation from the analytical model compared to Strategy H.

\begin{figure}[ht]
    \centering
    \includegraphics[scale=0.32]{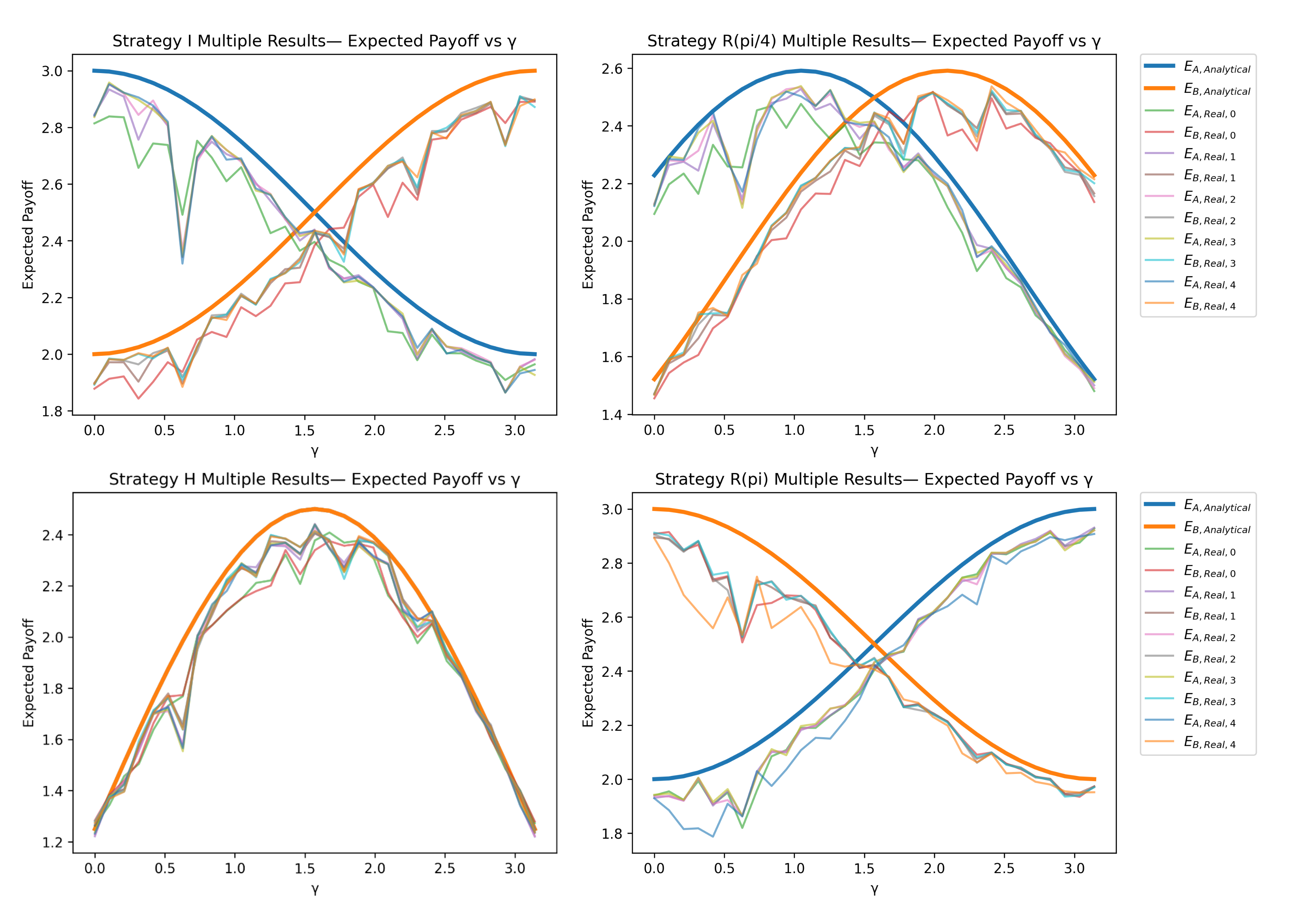}
    \caption{Multiple Real-Device Runs: Expected Payoff vs. $\gamma$ for Strategies I, H, $R\left(\frac{\pi}{4}\right)$, and $R(\pi)$}
    \label{fig:Fig_9}
\end{figure}

Figure~\ref{fig:Fig_9} shows the average payoffs from five real executions on \textit{ibm sherbrooke} for Strategies I (top) and H (bottom) as a function of the entanglement parameter $\gamma$, compared with the analytical curves $E_A$ and $E_B$. Each light-colored line represents one of the five real trials, showing the dispersion around the theoretical prediction and the overall consistency in the shape of the payoff distributions for both players.

In Fig.~\ref{fig:Fig_9} (top: strategy $R\left(\frac{\pi}{4}\right)$; bottom: strategy $R(\pi)$), the same dataset is presented: experimental payoffs for Alice and Bob for each $\gamma$ value across five repetitions, overlaid with the corresponding analytical curves. In both cases, the real series follow the morphology of the theoretical curves, with the $R\left(\frac{\pi}{4}\right)$ strategy centered on intermediate $\gamma$ values and the $R(\pi)$ strategy reflecting a trend inversion, while exhibiting pointwise variance due to hardware noise.

\subsection{Validation}

Table~\ref{tabla:recm} shows the average root mean square error (RMSE) from five runs for each player's payoff under each strategy. The highest RMSE is observed for Alice's payoff under Strategy I, and the lowest for Bob's payoff under Strategy $R\left(\frac{\pi}{4}\right)$.

\begin{table}[ht]
\centering
\caption{Values of $\sqrt{E_A\_RMSE}$ and $\sqrt{E_B\_RMSE}$ for different strategies.}
\begin{tabular}{|c|c|c|}
\hline
\textbf{Strategy} & \textbf{$\sqrt{E_A\_RMSE}$} & \textbf{$\sqrt{E_B\_RMSE}$} \\
\hline
H & 0.118 & 0.110 \\
\hline
I & 0.145 & 0.127 \\
\hline
R($\pi/4$) & 0.123 & 0.105 \\
\hline
R($\pi$) & 0.110 & 0.134 \\
\hline
\end{tabular}
\label{tabla:recm}
\end{table}

From Figures 5–8, it can be observed that the payoff scale ranges from a maximum of 3 to a minimum of 1.2. Based on this, a relative error percentage can be computed for the best and worst cases. These values are represented in Equations~(15) and (16), respectively.

\begin{equation}
    \epsilon_{+\%}=\frac{(0.105)(100)}{3}=3.5\%
\end{equation}

\begin{equation}
    \epsilon_{-\%}=\frac{(0.145)(100)}{1.2}=12.08\%
\end{equation}

The lowest error is observed for Bob under strategy $R\left(\frac{\pi}{4}\right)$ (RMSE = 0.105), while the highest error corresponds to Alice under strategy I (RMSE = 0.145). These results suggest that strategy $R\left(\frac{\pi}{4}\right)$ is more resilient to noise in the IBM hardware, particularly for Bob.

The relative error range of 3.5\% to 12.08\% indicates that although some deviation from the theoretical model exists, the observed payoffs generally follow the expected trends. This level of error is within acceptable limits for near-term quantum hardware, given the known sources of decoherence and gate infidelity.

The use of RMSE as a validation metric supports the conclusion that the Guided Circuit Mapping (GCM) strategy improves consistency, but further significance testing (e.g., paired t-tests) would be useful to confirm whether these observed differences are statistically meaningful across strategies.

\section{Discussion}

The experimental campaign consisted of only five repetitions per strategy and $\gamma$ value, a compromise imposed by limited runtime quotas on \textit{ibm sherbrooke}. Despite this modest sample size, clear systematic effects emerge. Noise in the expected payoffs grows sharply with qubit count and circuit depth: executing 31 two-qubit games (62 physical qubits) produces fluctuations that diverge from the analytical baseline, consistent with previous large-system observations~\cite{bluvstein2024logical}. Because the full Hilbert space is of dimension $2^{62}$, an exact analytic error model is intractable; the combination of gate infidelity, read-out error and crosstalk yields run-to-run variability that theory alone cannot capture.

The guided circuit-mapping strategy (GCM Strategy) attenuates that variability. Figures \ref{fig:juntas} and 5–8 show that GCM converts a visually chaotic scatter into curves that track the theoretical trend, even if confidence bands remain wide. The relative-error bounds derived from Table \ref{tabla:recm} $3.5\%$ (best case) to $12.1\%$ (worst case) quantify this residual gap. 

Variance matters because it reveals whether shot noise or hardware bias dominates the discrepancy.  With only five runs the standard error is still sizeable; expanding to $\ge 20$ runs would narrow confidence intervals by roughly a factor of 2, enabling a rigorous paired-$t$ significance test on each strategy.

Strategy-specific behaviour is also evident: Bob's payoff under $R(\pi/4)$ exhibits the lowest RMSE (0.105), hinting that certain local rotations align better with the device's calibration window, whereas Alice under Strategy I suffers the largest error (0.145).  Mapping-aware compilation could therefore be tailored to favour the most robust gate sets.

Upon analyzing Figure \ref{fig:Fig_9}, it becomes evident that, beyond some random fluctuations, the five runs on IBM Sherbrooke exhibit a defined and repetitive pattern within each strategy. This suggests a deterministic behavior in the results, likely caused by systematic hardware errors. A clear manifestation of this is the consistent dip or "valley" that appears across all strategies around $\gamma \!\approx\! 0.6$, a feature that is logically difficult to attribute to randomness. This deterministic pattern can be explained by considering that each quantum channel introduces independent systematic errors that affect the remaining channels. Without the GCM strategy, these errors interact in an uncontrolled manner, resulting in the chaotic curves observed in Figure \ref{fig:juntas}. In contrast, when the GCM strategy is applied, these systematic errors tend to average out at first order, thus smoothing their impact and contributing coherently to the observed value at each $\gamma$. In a forthcoming article, we will present a simplified analytical calculation that supports this hypothesis and provides further insight into the origin of these persistent structures in the experimental results.

Finally, the literature emphasises adaptive layout and zero-noise extrapolation as essential tools for NISQ fidelity \cite{bharti2022noisy}. Our results echo that view: even modest mitigation (dynamic qubit pairing) delivered measurable gains. Translating such advances to economics, logistics and mechanism-design use-cases will require both deeper statistical validation and continued progress in device-level error suppression~\cite{alodjants2024quantum,ikeda2022theory,ims2021quantum}.

\section{Conclusions}

Quantum strategies implemented under the Eisert-Wilkens-Lewenstein (EWL) framework demonstrate superior performance compared to classical BoS equilibrium, achieving up to 108\% improvement in theoretical scenarios and maintaining a 3.5\%--12\% advantage on real quantum hardware. Despite noise and systematic bias present on \textit{ibm sherbrooke}, the GCM Strategy successfully reduces RMSE to approximately 0.11--0.15 and establishes reproducible trends across five experimental executions, confirming the effectiveness of layout-aware compilation for NISQ devices.

The experimental design encountered statistical limitations due to the constraint of only five repetitions, which restricts formal hypothesis testing capabilities. Increasing the sample size to at least twenty runs would enable rigorous paired-$t$ tests and provide tighter confidence intervals for more robust statistical validation. Nevertheless, this study successfully demonstrates that the combination of layout-aware compilation and lightweight error-mitigation techniques can preserve quantum game payoffs within single-digit percentage error, representing a significant milestone for early NISQ applications.

The experimental results support the theoretical framework by showing that quantum error mitigation techniques effectively improve expectation values of observables on real quantum devices, while the observed persistence of quantum advantage under noisy conditions validates theoretical predictions regarding the robustness of quantum strategies in game-theoretic scenarios. These findings establish a practical foundation for quantum game theory implementation on current NISQ platforms and demonstrate the viability of quantum-enhanced coordination mechanisms in realistic hardware environments.

\section{Future Work}
Extended statistical validation represents a critical research direction that requires scaling experiments to at least 20--30 repetitions per strategy and $\gamma$ parameter to enable rigorous paired-$t$ tests and achieve tighter confidence intervals \cite{Joint_Prob}. This expanded statistical framework would clarify whether the observed performance differences between strategies demonstrate statistical significance and provide the robust validation protocols necessary for advancing quantum game theory from proof-of-concept demonstrations to practical implementation. Hardware-specific optimization should investigate the consistent performance degradation observed around $\gamma \approx 0.6$ across all strategies through comprehensive process tomography at specific entanglement values, real-time calibration data integration, and strategy-specific gate decomposition optimization tailored to individual quantum processor architectures \cite{9805433}.

Scaling investigations to multi-player games presents a natural extension from the current two-player focus, as recent research has demonstrated successful three-player quantum game implementations \cite{jaffali_two_2024} and four-qubit CHSH games \cite{jusseau2024fourqubitchshgames} that could evaluate the scalability of the GCM Strategy in more complex coordination scenarios. Error model development should prioritize the creation of simplified analytical models that capture first-order effects of systematic hardware errors under the GCM Strategy framework, providing theoretical foundation for observed error patterns and guiding future mitigation strategies \cite{fuchs_quantum_2020}. Cross-platform validation using different quantum processor technologies, including trapped ions and neutral atoms, would facilitate the distinction between hardware-specific artifacts and fundamental quantum game properties, following established protocols for cross-platform quantum state comparisons \cite{kelleher_implementing_2024,sheffer_playing_2022}.

The demonstrated quantum advantages, despite current error rates, indicate immediate potential for applications in supply chain coordination with quantum-secured communication, distributed resource allocation in smart grids, and multi-agent reinforcement learning with quantum-enhanced performance, building upon recent breakthroughs in quantum advantage for trading applications \cite{khan_quantum_2025}. These research directions would bridge the gap between theoretical quantum game advantages and practical implementations, ultimately enabling quantum-enhanced decision-making in complex multi-agent systems as NISQ hardware continues advancing toward fault-tolerant quantum computing \cite{bharti2022noisy,bluvstein2024logical}.

\section{Acknowledgment}
We thank the Universidad del Cauca and the Corporation for Aerospace Initiatives, Research and Innovation (CASIRI) for providing the facilities and working environment essential for the development of this research. We also extend our appreciation to IBM for free access to its quantum computing services, which were vital to the execution of the experiments presented here.

\section{Data and materials availability}
All source code required to implement the Guided Circuit Mapping Strategy (GCM Strategy), as well as to reproduce the numerical and experimental results, is publicly available on GitHub:\url{https://github.com/Carlosandp/GCMStrategy}.

\printbibliography

\end{document}